\documentclass[journal]{IEEEtran} 				
\IEEEoverridecommandlockouts 						
\usepackage{amsmath,amssymb}
\usepackage{amsthm}
\usepackage{color}
\usepackage[mathscr]{eucal}
\usepackage{graphics,graphicx,multicol}
\usepackage{epsfig}
\usepackage{enumerate}
\usepackage{subfigure}
\usepackage{algorithmic}
\usepackage{algorithm}
\usepackage{morefloats}
\usepackage{enumerate}
\usepackage{subfigure}
\usepackage{multirow}
\usepackage{array}
\usepackage{pstricks, pst-node, pst-plot, pst-circ}
\usepackage{moredefs}
\usepackage{cite}
\usepackage{courier}
\usepackage{epstopdf}
\usepackage{verbatim}

\begin{document}

\title{Position Estimation of Robotic Mobile Nodes in Wireless Testbed using GENI}
\author{
    \IEEEauthorblockN{Ahmed Abdelhadi\IEEEauthorrefmark{1}, Felipe Rechia\IEEEauthorrefmark{2},
					  Arvind Narayanan\IEEEauthorrefmark{3},
				          Thiago Teixeira\IEEEauthorrefmark{4},
					  Ricardo Lent\IEEEauthorrefmark{5},
					  Driss Benhaddou\IEEEauthorrefmark{5},
					  Hyunwoo Lee\IEEEauthorrefmark{6}, T. Charles Clancy\IEEEauthorrefmark{1}}\\
    \IEEEauthorblockA{\IEEEauthorrefmark{1}Virginia Tech (\{aabdelhadi, tcc\}@vt.edu)},
    \IEEEauthorblockA{\IEEEauthorrefmark{2}Arizona State University (feliperechia@asu.edu)},
    \IEEEauthorblockA{\IEEEauthorrefmark{3}University of Minnesota (arvind@cs.umn.edu)},
    \IEEEauthorblockA{\IEEEauthorrefmark{4}University of Massachusetts Amherst (tteixeira@umass.edu)},
    \IEEEauthorblockA{\IEEEauthorrefmark{5}University of Houston (rlent@uh.edu, dbenhadd@central.uh.edu)},
    \IEEEauthorblockA{\IEEEauthorrefmark{6}Seoul National University (hwlee2014@mmlab.snu.ac.kr)}\\
}

\maketitle

\begin{abstract}
We present a low complexity experimental RF-based indoor localization system based on the collection and processing of WiFi RSSI signals and processing using a RSS-based multi-lateration algorithm to determine a robotic mobile node's location. We use a real indoor wireless testbed called w-iLab.t that is deployed in Zwijnaarde, Ghent, Belgium. One of the unique attributes of this testbed is that it provides tools and interfaces using Global Environment for Network Innovations (GENI) project to easily create reproducible wireless network experiments in a controlled environment. We provide a low complexity algorithm to estimate the location of the mobile robots in the indoor environment. In addition, we provide a comparison between some of our collected measurements with their corresponding location estimation and the actual robot location. The comparison shows an accuracy between 0.65 and 5 meters. 

\end{abstract}

\pagenumbering{gobble}

\section{Introduction}\label{sec:intro}

Wireless technologies are enabling numerous applications impacting positively the quality of our lives. They are playing an important role in reducing the digital divide of underserved population and providing value added services such as ubiquitous access. They play a significant role in location based services (LBS). In particular positioning users within an area plays a key role in location based services, such as emergency, medical, delivery system, supermarket, parking lots, restaurant, security and manufacturing applications. End users can benefit from location-aware services, such as printing to the nearest printer or navigating through a building \cite{Bahl_Padmanabhan_2000}. Wireless network administrators can use client location in a network to speed up troubleshooting and also to detect rogue clients and access points \cite{cisco_wifi_lbs}.

Global Positioning System (GPS) is a mature technology and widely used in outdoor location based services \cite{misra1999special}. However, GPS does not provide accurate information indoors. Currently several different indoor location technologies are being investigated. Technologies include mobile communication system, Wireless Sensor Networks (WSN), and WiFi networks. Mobile communication systems have poor precision due to the low density of deployed base stations, while WSN based positioning systems would require additional sensors carried by the user or embedded in the mobile phone, and thus would not be a very practical and cost effective approach.

WiFi-based positioning systems are widely considered as a good candidates as access points are already deployed in high density in cities, university campuses, and homes. Using WiFi, location estimation systems can provide indoor positioning information with a precision of a few meters \cite{Bahl_Padmanabhan_2000}, \cite{Brown2011}. Current location systems developed and in use so far use one or more techniques to achieve this goal.

Considering a system composed of a mobile node and several fixed access points, some of the most common location techniques are briefly summarized here \cite{Brown2011}:
\begin{itemize}
	\item \textbf{Fingerprinting}: A database of the RSSI values across a location is recorded in a calibration phase (fingerprinting) by a mobile node. Then in the online phase, the location of the device is determined by matching the currently measured RSSI values to the database.
	\item \textbf{Lateration}: Using 3 or more reference access points, tri- or multi-lateration is possible by estimating their distance to the mobile node. The position of the mobile node can be calculated using a function that is composed of the lengths between the access points detected and the mobile device \cite{Brown2011}.
	\item \textbf{Angulation}: Using 2 or more reference access points, this technique computes angles between access points and the mobile node to determine its position.
\end{itemize}

Current commercial wireless network design guides suggest that the combination of these techniques is usually done to increase the overall location estimation accuracy \cite{cisco_wifi_lbs}.

In this paper we present an experimental localization system based on the collection of WiFi RSSI signals and processing using a multi-lateration algorithm to determine the mobile node's location. We use a testbed called w-iLab.t and deployed in Zwijnaarde, Belgium. One of the unique attributes of this testbed is that it provides tools and interfaces to easily create reproducible wireless network experiments in a controlled environment. W-iLab.t offers fixed and mobile wireless nodes, all of which may be controlled entirely remotely \cite{becue2012}. Additionally, this testbed is federated to larger testbed organizations such as Federation for Future Internet Research and Experimentation (Fed4Fire\footnote{``Fed4FIRE is an Integrating Project under the European Union’s Seventh Framework Programme (FP7) addressing the work programme topic Future Internet Research and Experimentation'' \cite{fed4fire}.}) and GENI\footnote{``GENI is a new, nationwide suite of infrastructure supporting "at scale" research in 
networking, distributed systems, security, and novel applications. It is supported by the National Science Foundation, and available without charge for research and classroom use'' \cite{geni}.}, meaning that it can be used by researchers from virtually anywhere to easily create and reproduce network experiments. Additionally, the outcome of location estimation can benefit other systems when integrated, e.g. ad-hoc networks \cite{Jose_INFOCOM2010, Abdelhadi_ITW2010}, networks power allocation \cite{AbdelhadiarXiv2014_1, WangarXiv2015_1, WangarXiv2015_2} and HetNets resource allocation \cite{Shajaiaharxiv2015_1, AbdelhadiICNC2015}.

\subsection{Related Work}\label{sec:related}

In the past decades many projects demonstrated ways of determining the position of mobile nodes within a wireless network using a variety of methods such as IR, acoustic ultrasound, proprietary microwave, RFID, bluetooth, cellular networks,  ZigBee, WiFi and UWB \cite{Yang2013}, \cite{Brown2011}. In general these methods are based on either fingerprinting or triangulation \cite{Brown2011}. Indoor localization based on wireless networks is even present in commercial solutions such as CISCO's WiFi LBS \cite{cisco_wifi_lbs}.

The RADAR solution by Bahl et al. \cite{Bahl_Padmanabhan_2000} was probably one of the first attempts to determine the user's location based on RSSI values collected from broadcast packets called \textit{beacons}. Their solution used the mobile node to send the beacons while the fixed access points simply measured the signal strength. Our experiment uses the opposite approach, having the mobile node listening for beacons coming from multiple different access points while we measure the signal strength in the mobile node. Chandrasekaran et al. \cite{ChandrasekaranEYLCGM09} empirically evaluates the accuracy limits of RSS based localization techniques, which gives us a reference as to what kind of results we should expect.

Redpin \cite{redpin} is an open source project which provides room-level accuracy indoor positioning service with zero-configuration, which means that it omits the calibration phase. The method employed by Redpin uses signal-strength of GSM, Bluetooth and WiFi access points on a mobile phone to estimate the location of the device \cite{Bolliger:2008:RAZ:1410012.1410025}. OpenBeacon \cite{openbeacon} is a project using active low power RFID devices which work as beacons and specialized USBnode base stations connected to mobile nodes to read beacon signals. The base stations process beacon packets received from stationery beacons and compute their own position relative to the fixed beacon devices.

CISCO Location-Based Services architecture provides a comprehensive indoor positioning system based on reading RSSI from three or more fixed access points and using a RSS tri- or multi-lateration approach to calculate the client's position. Their method uses a calibration phase in which they sample and record radio signal behavior to create a \textit{radio map} of the entire location (e.g. campus, office). The data acquired during the calibration phase is then used to better estimate the client's position during the operational phase of the system \cite{cisco_wifi_lbs}.

Abdel-Hadi et al. \cite{Ahmed_VTC10} presented a simulation model for Horus testbed, a network of autonomous aerial vehicles (AAV)s where the AAVs communicate wirelessly. They determine the quality of channel for video transmission using RSSI. The same method was implemented in a real-testbed in \cite{Ahmed_Horus_project}.

Zhang \cite{ Zhang2011} proposed an indoor location technique using WiFi wireless signals utilizing dead reckoning method that uses step number and step length related to a random motion of the mobile device. The technique improves the precision by using WiFi signal auxiliary positioning.

CAMMEO \cite{ loreti2012} system has been proposed to implement a positioning system that integrates heterogeneous technologies such as WiFi and Bluetooth. It implements interfaces in distributed system that use radio frequency, the positioning algorithms, and the applications. The purpose of the system is to provide direct access to wireless technologies, integrate existing localization framework, and make standard interface available for developers. A prototype CAMMEO system was developed and tested.

Chong \cite{Chong2013} proposed a technique based on probability-weighted algorithm that computes the probability of the pre-positioning point to complete the positioning simultaneously. The algorithm improves the response time and precision of the positioning.

Zirari et al. \cite{zirari2013} proposed a hybrid system that utilizes an existing positioning system and uses 802.11 to improve the precision of the positioning system. The algorithm uses assessment and dilution technique to improve the precision.

Van Haute et al. \cite{vanhaute15_platform} present the EVARILOS Benchmarking Platform (EBP), which aims to automatically evaluate multiple indoor localization solutions, even if these solutions were developed in different environments and with different evaluation metrics. EBP even allows experimenters to analyze public datasets of previously collected RF data to test their own solutions.

\subsection{Our Contributions}\label{sec:contributions}

Our contributions in this paper are summarized as:
\begin{itemize}
	\item We demonstrate the use of an innovative testbed \cite{becue2012} that uses GENI project and allowed us to prototype our experiment using mobile robots with WiFi interfaces. This testbed allows the experiment to be reproduced and modified for future enhancements as well.
	\item We show the use of a low complexity RSS-based lateration algorithm which allows us to estimate the position of the robot and compare it to the real position given by the testbed framework.
\end{itemize}

The remainder of this paper is organized as follows. Section \ref{sec:Testbed} presents the testbed setup and some details about its operation. In Section \ref{sec:Algorithm}, we present our robot location estimation algorithm. Section \ref{sec:Performance and Evaluation} provides quantitative evaluation results along with discussion. Section \ref{sec:conclude} concludes the paper and discusses the possibility of future work.

\section{Testbed Description}\label{sec:Testbed}

\begin{figure*}[t]
\centering
  \includegraphics[width=\linewidth]{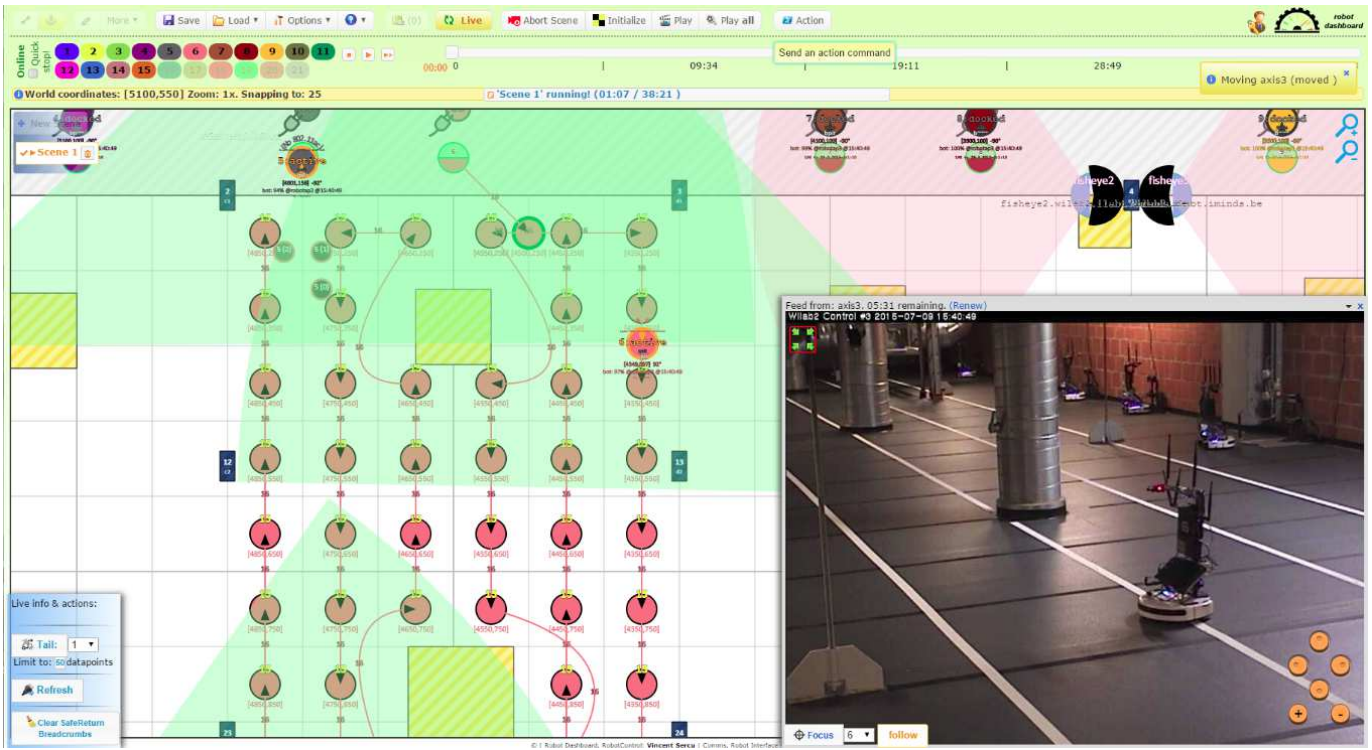}
  \caption{Screenshot of w-iLab.t's robot control front-end.}
  \label{fig:screenshot}
\end{figure*}

The experiments were carried out on \textit{w-iLab.t}--a Generic Wireless Testbed hosted by iMinds\footnote{``iMinds is Flanders' digital research center located in Zwijnaarde, Ghent, Belgium. It includes 900+ researchers at 5 Flemish universities. It conducts strategic and applied research in areas such as ICT, Media and Health'' \cite{iMinds}.} and located in Zwijnaarde, Belgium and that offers remote access to stationary and mobile network equipment. The testbed nodes are Linux-based PCs equipped with IEEE 802.11a/b/g/n, IEEE 802.15.4, and IEEE 802.15.1 interfaces. To achieve mobility, a subset of the PCs have been installed on top of Roomba vacuum cleaning robots, which are controlled by a central testbed manager through a wireless interface to the Roomba Open Interface (ROI) \cite{becue2012}. The interface allows to override most of the normal Roomba functions and to define a custom mobility, which can be done online through the testbed manager front-end, either visually of programmatically, by entering 
the individual trajectory and moving speed for each of the robots. By having separate and unrestricted access to the PCs mounted on the robots, the testbed offers a flexible platform for experimentation with mobile communications. For example, the power transmission of each radio can be defined by using the wireless tools package for Linux. The robots lack dedicated localization hardware. However, the testbed manager can estimate the location of each robot with approximately 1 cm precision error, by measuring the distance traveled from their docking station. The estimations are regularly adjusted by the system as the robots cross markers that are permanently attached to the floor, which can be sensed by the robots \cite{becue2012}. The estimated location of the robots can be queried via a REST interface, and were used to evaluate the performance of the localization algorithms that we tested. A screenshot of the robot control interface (or the Robot Dashboard) is shown in Figure \ref{fig:screenshot}. The 
robot control front-end provided an interface to design and control the robot movements. It also provided access to real-time visuals of the testbed. The red circles in Figure \ref{fig:screenshot} shows the user-assigned paths to a particular Roomba robot, such that when the experiment is run the   robot will follow this path. The yellow rectangular objects show the physical obstacles on the testbed floor. The path we define should avoid such obstacles. Green pie-shaped portions indicate the visibility from the camera located at every active robot on the testbed. The testbed also consists some fixed fish-eye and standard cameras. For example, the bottom-right in Figure \ref{fig:screenshot} show live visuals of the testbed as seen from one of the fixed cameras.

\subsection{Experiment Setup}\label{sec:Testbed:B}

The w-iLab.t iMinds testbed is part of the Fed4Fire project--a large federation of testbeds in Europe, which also allows GENI users to gain access through the use of GENI portal credentials \cite{geni}. Similar to any GENI experiment, first we defined the topology and assigned the resources required for the tests using jFed--a Java-based tool that allows to visually create, manage and run experiments. This tool generated a RSpec file which contained all the specifications required to reserve particular equipment multiple times for several experiments.

Once the resources were reserved, we used the robot control front-end to assign the path for one robot, and made it move to a number of locations. The fixed access points were simply configured to host SSIDs corresponding to their hostnames and using a variety of 802.11a channels. A Python script was made to run at this mobile node. As the mobile node (or robot) moved from one place to another, the script was  configured to periodically scan the network, and record the RSSI values (by calling \textit{iwlist}) associated with every access point. Our RSpec file and other scripts written to setup this experiment are stored in a GitHub repository. 

\section{Location Estimation Algorithm}\label{sec:Algorithm}
In this section, we present a low complexity algorithm for estimating the location of the robot in the testbed described in Section \ref{sec:Testbed}. We divide the algorithm into three stages: initial calibration, power measurement for new location, and coordinates calculation stages. Note that this algorithm could be generalized for any testbed.\\
\textbf{Initialization (Calibration stage):} in this stage we calculate path loss exponent $\alpha$ of the medium.

    \begin{itemize}
      \item Measure power at the fixed nodes in dBm for a robot location with know coordinates $(x_0,y_0)$, e.g. docking location of robot in the testbed. Record the highest $M$ powers received from fixed nodes, e.g. let $M=4$ we measure $P_1,P_2,P_3,P_4$.
      \item Calculate distance from these $M$ fixed nodes, e.g. $r_1,r_2,r_3,r_4$. Store these power and distance values in a database.
      \item Use power received from any two fixed nodes of the $M$ fixed nodes $P_i$ and $P_j$ and distance $r_i$ and $r_j$ to calculate $\alpha$  using
      \begin{equation}
      \alpha_l = \frac{P_i-P_j}{10\log(\frac{r_j}{r_i})}.
      \end{equation}
      \item Repeat path loss exponent calculation $\binom{M}{2}$ times (i.e for each two fixed nodes combination). Then average $\alpha$ values to get an estimate
      \begin{equation}
	  \hat{\alpha}= \frac{\sum_{l=1}^{\binom{M}{2}} \alpha_l }{\binom{M}{2}}.
      \end{equation}
      \item Repeat calibration for other known locations if possible to have a better estimate of $\alpha$.
      \end{itemize}
\textbf{Power Measurement for Unknown Robot Location:}

  \begin{itemize}
  \item Measure highest $N$ fixed nodes received power $P_k$ for $k=\{1,2,...,N\}$ and compare with power and distance measured from known locations $P_i$ and $r_i$ respectively (i.e. use the values stored in the database in the calibration stage) using the following equation
  \begin{equation}
   r_k = r_i 10^{(\frac{P_i-P_k}{10\hat{\alpha}})}.
  \end{equation}
  \item Repeat for all values for power and distance stored database. Then average the values to get $\hat{r}_k$ using
        \begin{equation}
	  \hat{r}_k= \frac{\sum_{l=1}^{\binom{M}{2}} {r}_k^l }{\binom{M}{2}}.
      \end{equation}
  \item Repeat the previous two steps for all $N$ fixed nodes.
  \end{itemize}
\noindent
\textbf{Coordinates Calculation:} In this stage, we construct two circles with centers at two of the $N$ fixed nodes and radii equal the corresponding estimated distances.  We used the following procedure to calculate the intersection point between the two circles:

 \begin{itemize}
 \item For fixed nodes $i$ and $j$ with coordinates $(x_i,y_i)$ and $(x_j,y_j)$ and estimated distances $\hat{r}_i$ and $\hat{r}_j$ respectively, use the following equations to calculate intersection points $(x_0,y_0)$
\begin{align*}
d&=\sqrt{(x_i-x_j)^2+(y_i-y_j)^2}\\
l&=\frac{\hat{r}_i^2-\hat{r}_j^2+d^2}{2d},\quad h=\sqrt{\hat{r}_i^2-l^2}\\
x_0&=\frac{l}{d}(x_j-x_i) \pm \frac{h}{d}(y_j-y_i) + x_i,\\
y_0&=\frac{l}{d}(y_j-y_i) \mp \frac{h}{d}(x_j-x_i) + y_i.
\end{align*}
\item Since the equations can produce up to two solutions for the intersection point,
the selected point is the one closer to the other $N$ fixed nodes. In the case of no intersection, we use the
mid-point between the two centers. Repeat the coordinate calculations $\binom{N}{2}$ times.

\item Average the coordinates for an estimate of the location $(\hat{x}_0, \hat{y}_0)$.
\end{itemize}

\section{Performance and Evaluation}\label{sec:Performance and Evaluation}
After the implementation of the aforementioned algorithm, we ran our tests using one robot and a set of 36 fixed access points. Our scripts ran every minute, collecting data and parsing it to a MySQL server, generating a database of APs ordered by RSSI, SSID, and robot location.
The following subsections describe the results encountered -- where we generated a heat map to help visualize the signal strength -- and the location calculated compared with the actual location of the robot.

\subsection{Initial Observations}\label{sec:heat}

\begin{figure}
\centering
\includegraphics[width=\linewidth]{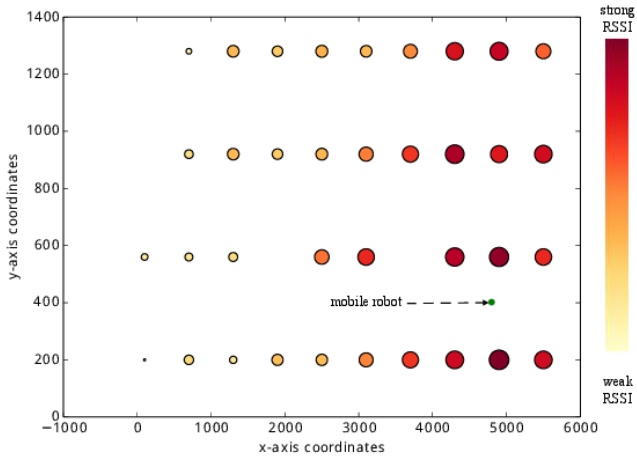}
\caption{Bubble map showing the average RSSI value seen by the bot with respect to the fixed wireless nodes in a 2-D spatial coordinate system}
\label{fig:bubblemap}
\end{figure}

The data collected from the bot was translated into a bubble map, shown on Figure~\ref{fig:bubblemap}. Both the \textit{x-axis} and \textit{y-axis} represent the geo-coordinates of the fixed access point and the robot in a 2-D space. The scale of the coordinates are in centimeters. The green marker represents the location of the mobile bot (or robot), while the other bubbles represent the location of the fixed access points. The size of these bubbles are proportional to the average RSSI value observed by the mobile robot with respect to the fixed access points. The intensity is also translated into a color code, represented in the right color palette.

As expected, the access points close to the robot express higher signal strength and the ones farther from the robot show lower signal strength. Some discrepancies may occur due to the nature of the testbed, such as reflections of the air conduits, attenuation or even blocking of the building columns.

\subsection{Results and Location Estimation Error}\label{sec:error}
During our experiments, we navigate the robot to several positions in the testbed, collect data, and perform the calculations to estimate its position. Some of the points are shown in Table I and are ordered by error. As we can see, we achieved an accuracy error close to one meter in some cases. We believe that these measurements were done at what we define as good spots (with minimal objects reflecting signals and creating multiple paths). In some of the cases, we achieved an accuracy of a few meters, which still considerably good in an environment like our testbed.

\begin{table}[]
\centering
\caption{Position estimation error}
\label{my-label}
\begin{tabular}{llll}
\hline
Estimated position & Actual Position & Error (in cm)\\
\hline
(4339.1, 591.54)		& (4399, 574) & 61.4	\\
(4971.58, 476.27)		& (4899, 398) & 106.15	\\
(4790.35, 692.79)		& (4899, 305) & 401.79	\\
(4580.14, 878.62)		& (4899, 502) & 492.44	\\
\hline
\end{tabular}
\end{table}

\section{Conclusion}\label{sec:conclude}

In this paper, we implemented a low complexity RSS-based lateration algorithm for localization in WiFi networks using GENI. Our results show an accuracy between 0.65 and 5 meters between the actual and estimated location of the mobile node. In addition, we demonstrated a use case of the mobile w-iLab.t robot testbed. We were able to setup and run this experiment smoothly using the provided interfaces tools of the w-iLab.t testbed and GENI project tools. This work demonstrates the importance of the availability of such testbeds and tools, where researchers can quickly prototype and develop solutions for a great number of use cases, including but not limited to indoor positioning systems.

\subsection{Future Work}\label{sec:futurework}

While analyzing our results, for some cases the mobile robotic node would not detect some of the SSIDs present in the testbed or give inaccurate RSSI measurements. We attribute this results to multiple paths and reflection of air conduits. Multiple paths are a well-known wireless problem and we believe its effects are amplified by the indoor test environment. This is expected, as most of the literature points out that RSSI measurements tend to fluctuate and even disappear \cite{ChandrasekaranEYLCGM09}. In future work, we plan to consider using CSI based localization techniques to enhance our algorithm for a more robust estimation.

Currently our application does not run in real time. Our experimental application collects the signal strength information from several locations of the testbed during a test run, and the processing of the information is performed offline. We could improve our application to provide real time localization results of the mobile node.

Our algorithm uses a single $\alpha$ value for path loss exponent of the indoor environment. It is possible to improve the accuracy of our positioning system by considering a variable path loss exponent which depends on the position, having or not having line-of-sight, and other medium characteristics which could also be recorded in a database. Thus our system would combine RF fingerprinting with RSS-based lateration techniques.

User orientation (north, south, etc) might also be a valuable piece of information considered in location estimation. In a real system, it could also be shown to be informative to determine if the user's body is actually obstructing the wireless signal \cite{Bahl_Padmanabhan_2000}. Mobile devices such as smartphones usually are equipped with a magnetometer which can be used to provide this additional information. The w-iLab.t facility provides enough flexibility so that smartphones could be attached to the mobile nodes and used in the experiments.

In the future, we could also evaluate our system using the Evarilos Benchmarking Platform proposed by Van Haute et al. \cite{vanhaute15_platform}. This would help us understand how accurate our positioning system is in comparison to others.

\section*{Acknowledgment}
We would like to thank iMinds researchers for their help with the experiment setup, especially Brecht Vermeulen, Vincent Sercu, and Pieter Becue,
and the GENI Project Office for partially supporting this work.

\bibliographystyle{ieeetr}
\bibliography{pubs}
\end{document}